\documentclass[letterpaper,twocolumn,showpacs,pra,aps]{revtex4}
\usepackage{amsmath}  
\usepackage{amsfonts} 
\usepackage{graphicx} 
\usepackage{amssymb}

\begin{document} 
\title{Improved prescription for winding an electromagnet}
\author{Christopher B. Crawford and Joseph P. Straley}
\affiliation{Department of Physics and Astronomy, University of Kentucky, Lexington, KY 40506-0055, USA}
\date{July 4, 2020}
\begin{abstract}
We describe an improvement on the magnetic scalar potential approach to the design of an electromagnet, which incorporates the need to wind the coil as a helix. Any magnetic field that can be described by a magnetic scalar potential is produced with high fidelity within a Target region; all fields are confined within a larger Return.  The helical winding only affects the field in the Return. 
\end{abstract}
\maketitle

\section{Using the magnetic scalar potential to design a coil}

In practical uses of magnetism it is sometimes desirable to be able to create a very well characterized magnetic field; for example, a field that is uniform within a defined region (hereafter, the Target), and with a surrounding region (the Return) which confines the field, so that there is no magnetic disturbance outside the device.     Ref~\cite{CBC2020} has described an algorithm to do this that makes use of the magnetic scalar potential to determine the surface current densities  on the interfaces. 

Here is how the algorithm works:   the specified magnetic field inside the Target is represented by a scalar potential such that $\vec H_{target} (\vec r) = - \vec \nabla \Phi_{target} $.  
A current field $\vec K(\vec r)$ at the surface of the Target is constructed so that only the normal component  
of $\vec H$ is present just outside the target.   It can be shown that this surface current density flows along the equipotentials of $\Phi_{target}$; having the current between any two equipotentials be equal to the potential difference ensures that the tangential components of the field are terminated.  This establishes the physical interpretation of magnetic scalar potential as a ``source potential'' in analogy with charge being the source of electric flux lines.

The Return envelopes the Target, excepting possibly parts of the Target where there is no normal field.  The normal component of the field is already specified at the interface between the Target and the Return, and it is required to be zero at the exterior surface.   Then the magnetic scalar potential $\Phi_{return}$ defined in the Return is determined by Neumann boundary conditions.   The current field on both the inner and outer surfaces of the Return is determined by this scalar potential in the same way as above.   

The result is a complete description how to construct surface current distributions that exactly produce within the Target  any field configuration that is consistent with Maxwell’s equations.   Choosing equal increments of the scalar potentials divides the surfaces into ribbons which can be turned into physical wires, all carrying the same current.

For the cases that the Return is one region that completely envelopes the Target, a  simplification of this algorithm is possible: as described, there are two current sheets right next together at the interface between the Target and the Return, and these can be merged.  Then the current sheet on the surface is determined by the gradient of the difference between the two scalar potentials, and flows along contours of constant value of this difference.   In the continuum limit (i.e. very fine wires) the surface current density is 
\begin{equation}
\label{surfacecurrent}
\vec K (\vec r) = - \hat n \times \vec \nabla \delta \Phi  ,   
\end{equation}
where the stream function~\cite{Haus, Mayergoyz, Lemdiasov} $\delta \Phi = \Phi_{return} - \Phi_{target}$ is the difference between the scalar potentials on the two sides of the surface and $\hat n$ is the outward-directed normal.

A small source of dissatisfaction with this algorithm for constructing designed-field coils is that what it prescribes is the current that should flow in closed loops around the surface of the Target, and on the surfaces of the Return.    The effect of breaking each loop and connecting it to its neighbors to make a series circuit introduces the equivalent of a single wire lying transverse to the loops; its effect can be nearly canceled by running the return current wire right on top of the interconnections, but this leaves a magnetic dipole line source.   Since the perfect field envisioned by the algorithm is otherwise only spoiled by exponentially small corrections due to wire discreteness, this would appear to be the largest design error in the constructed field.

Here we will describe a  modification of the algorithm which removes this defect, allowing the construction of a more nearly perfect coil.  In the next section we will work though a problem is that exactly solvable, but not quite trivial; later we will generalize this to arbitrary geometry.

\section{Spherical electromagnet}

Winding an infinite solenoid as a helix also introduces an axial current, but this does not affect the field inside it.  This can be readily generalized to any azimuthally symmetric object.  First consider the case that the field in the Target is zero, but there is with an axial current on the surface of the Target. Let the symmetry axis be $z$, and the shape of the object be $\rho(z)$.  Current conservation requires that the current through any cross section be the same; then from outside this is indistinguishable from a long straight wire along the axis (this wire will have to actually exist beyond the object); the added field outside will be the corresponding field, which has only a $\hat \phi$ component. This adds no field component normal to the surface, and the parallel component is exactly canceled by the surface current.  The field inside is unaffected.  Adding the axially symmetric field inside the Target, we find the corresponding field in the Return by the algorithm described above, and add the currents and fields just constructed to find a consistent set of fields produced by wires that wind around the surfaces.  

For the case of a sphere of radius $P$ with only an axial current, the surface current density is $K_{axial} = - \hat \theta I/(2 \pi P \sin \theta)$  (in spherical coordinates) and the field outside is  $\vec H =   \hat \phi I/2 \pi r \sin \theta$.   We can represent this as a magnetic scalar potential $\vec H = - \vec \nabla \Phi_{axial}$,  where 
\begin{equation}
\label{axial}
\Phi _ {axial} = -\phi I /2 \pi
\end{equation}
This is a multiply valued function; we can make it single valued by introducing a branch surface interrupting any path that wraps around the sphere or the wire extension. It has the same discontinuity $I$ everywhere across the branch surface, which can be taken to be the half-plane $\phi = 0$ for $r > P$ -- but note that the gradient of $\Phi_{axial}$ (the magnetic field) is continuous everywhere except at the poles $\theta = 0$ and $\theta = \pi $.

Now consider how the magnetic potential construction describes the same sphere when it has a uniform magnetic field inside (but no axial current), and all magnetic flux is enclosed within a Return of radius $Q$.  Inside, the magnetic potential is   
\begin{equation}
\label{target}
\Phi_{target} = -zH_0 = -H_{0} r \cos \theta   .
\end{equation} 
The potential in the Return is
\begin{equation}
\label{return}
\Phi_{return} = H_{0}  \frac {P^3 (r + Q^3/2r^2) \cos \theta}{Q^3-P^3}.
\end{equation}

At the boundary the normal component of the magnetic field is continuous, so that $- \frac {\partial} {\partial r} \Phi_{target} = 0$.   The surface current density that matches up the tangential magnetic fields is 
\begin{equation} 
\vec K_{inner} =  \hat \phi \frac{3}{2} H_{0} \frac {Q^3}
{Q^3-P^3}\sin \theta \equiv \hat \phi S_{inner} \sin \theta  
\end{equation}
on interface between the Target and the Return, and
\begin{equation}
\vec K_{outer} = -\hat \phi \frac{3}{2} H_0 \frac{P^3}{Q^3-P^3} \sin \theta \equiv - \hat \phi S_{outer} \sin \theta 
\end{equation}
on the outer surface of the Return.
If we slice up the sphere surface at equal intervals of the magnetic potential, or equivalently at equal intervals of the coordinate $z$ (call this interval $D$), each of the resulting rings on the surface of the Target is carrying the same current $\vec I =  \hat \phi D S_{inner}$. 
The rings have the same extent in the coordinate $z$ but varying width on the surface of the sphere.

Now we make this set of rings helical, so that after one turn the bottom of this ring is higher (measured along the $z$ axis) by the amount $D$.   This adds an axial current 
$\vec I_{helix} = \hat z D S_{inner}$ in 
the form of the surface current density $K_{axial}$. This modifies the potential in the Return $\Phi_{return}$ (\ref{return}) by adding an axial potential of the form $\Phi_{axial}$ (\ref{axial}) given above.  As explained above, this has no effect inside the sphere.

\begin{figure}[tb]
\label{fig:helix}
\includegraphics[width=1.0\columnwidth,keepaspectratio]{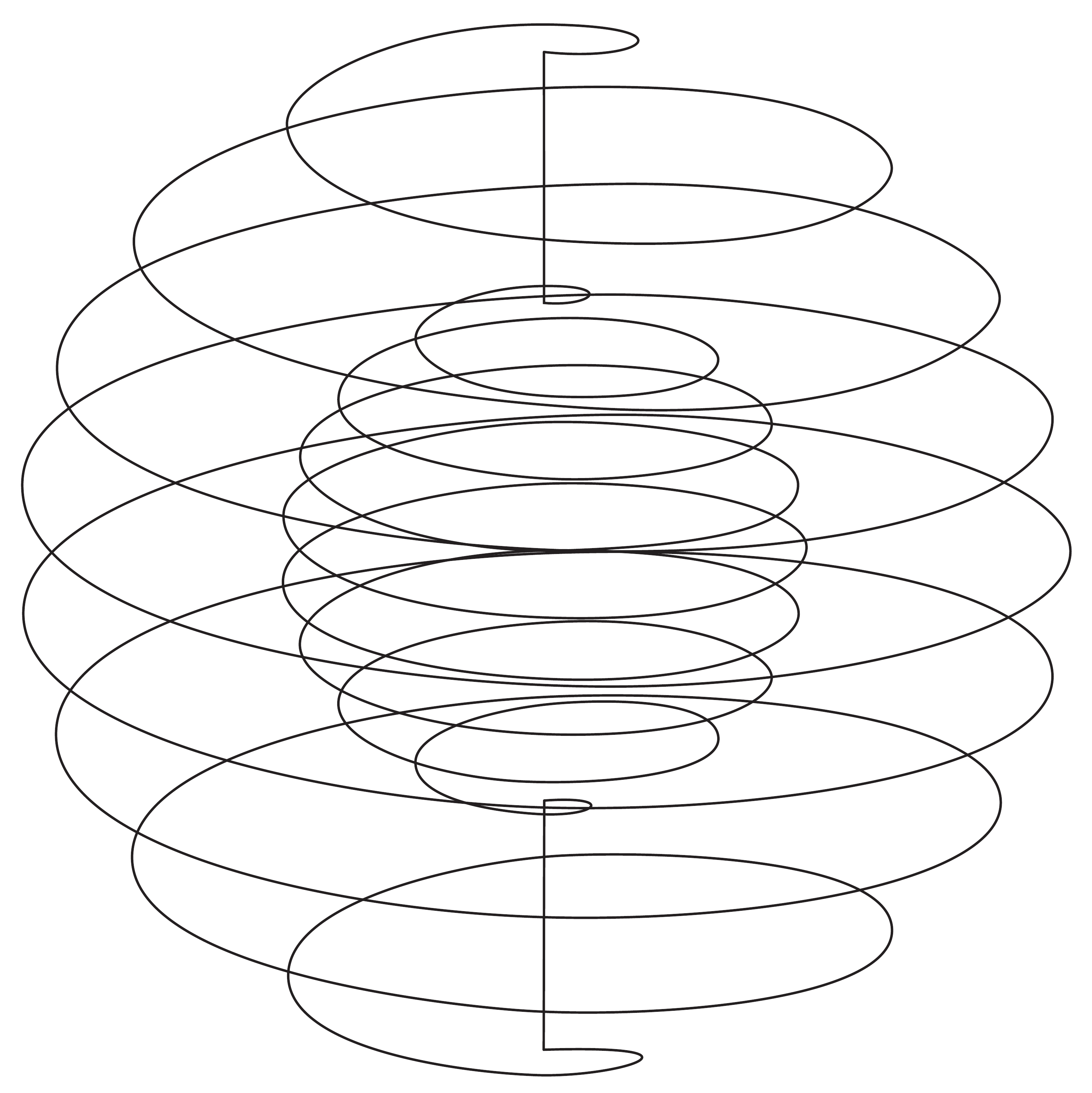}
\caption { An illustration of the helical winding for the spherical electromagnet.  An actual coil would use a much smaller current and far more turns}
\end{figure}

The interesting point is that in moving up one turn, the scalar potential difference $\Phi_{return} - \Phi_{target}$ decreases by $D S$,  and in going around one turn, the potential outside due to the axial current $\Phi_{axial} $ increases by $D S$.
This shows that the condition $\Phi_{return} + \Phi_{axial}-\Phi_{target} = constant$ generates a helical winding for a spherical electromagnet that completely and efficiently covers the surface.  The same construction applies to the outer surface of the return with the omission of $\Phi_{target}$; observe that all of the current that was moving along the $z$ axis on the Target will return along the outer surface in agreement with both Ampere's law and current conservation.

 Including the effects of the axial currents, the magnetic potential in the Return is given by 
\begin{equation}
\tilde \Phi_{return} = H_0  \frac {P^3 (r + Q^3/2r^2) \cos(\theta)}{Q^3-P^3}   - I \frac {1}{2 \pi} \phi
\end{equation}
The potential difference across the surface of the target and across the outer surface of the return determines the surface current density.  Slicing each boundary surface along lines of constant potential difference in increments of $I$ turns each one into a ribbon that winds around the surface, connecting up with the next ribbon. Explicitly, the potential differences are
\begin{equation}
\delta \tilde\Phi_{inner} = \frac {3}{2} H_0 P \frac { Q^3}{Q^3-P^3}\cos \theta - I \frac{1}{2 \pi} \phi
\end{equation}
at the interface between the Target and the Return, and
\begin{equation}
\delta \tilde \Phi_{outer} =  - \frac {3}{2}  H_0 Q \frac { P^3}{Q^3-P^3}\cos \theta + I \frac{1}{2 \pi} \phi
\end{equation}
on the exterior surface of the Return.
The tangential (along the surface) gradient of the potential differences corresponds to a surface current density (Eq. \ref{surfacecurrent}) that twists about the sphere in such a way as to match the tangential component of the field inside with the tangential field outside according to Ampere’s law. Beyond the Target, there are a pair of wires along the axis, carrying the current $I$ to the outer surface of the Return. In constructing the spherical magnet, the field inside and the current $I$ are independent parameters, which determine the width of the windings (or the spacing of the equivalent wires).  Once the potentials have been chosen, the path of the windings is determined by the condition of constant potential difference. 

The combined potential $\tilde\Phi_{return}$ actually has a nonphysical discontinuity $DS$ at each crossing of the branch cut, which must be built into numerical solutions, but the result is equivalent to the multi-valued $\Phi_{axial}$ potential with no discontinuities as describe above, for which the entire winding is traced out by a single equipotential contour on both the inside and outside of the Return. 

\section{the general case}

\begin{figure}[tbh]
\includegraphics[width=1.0\columnwidth,keepaspectratio]{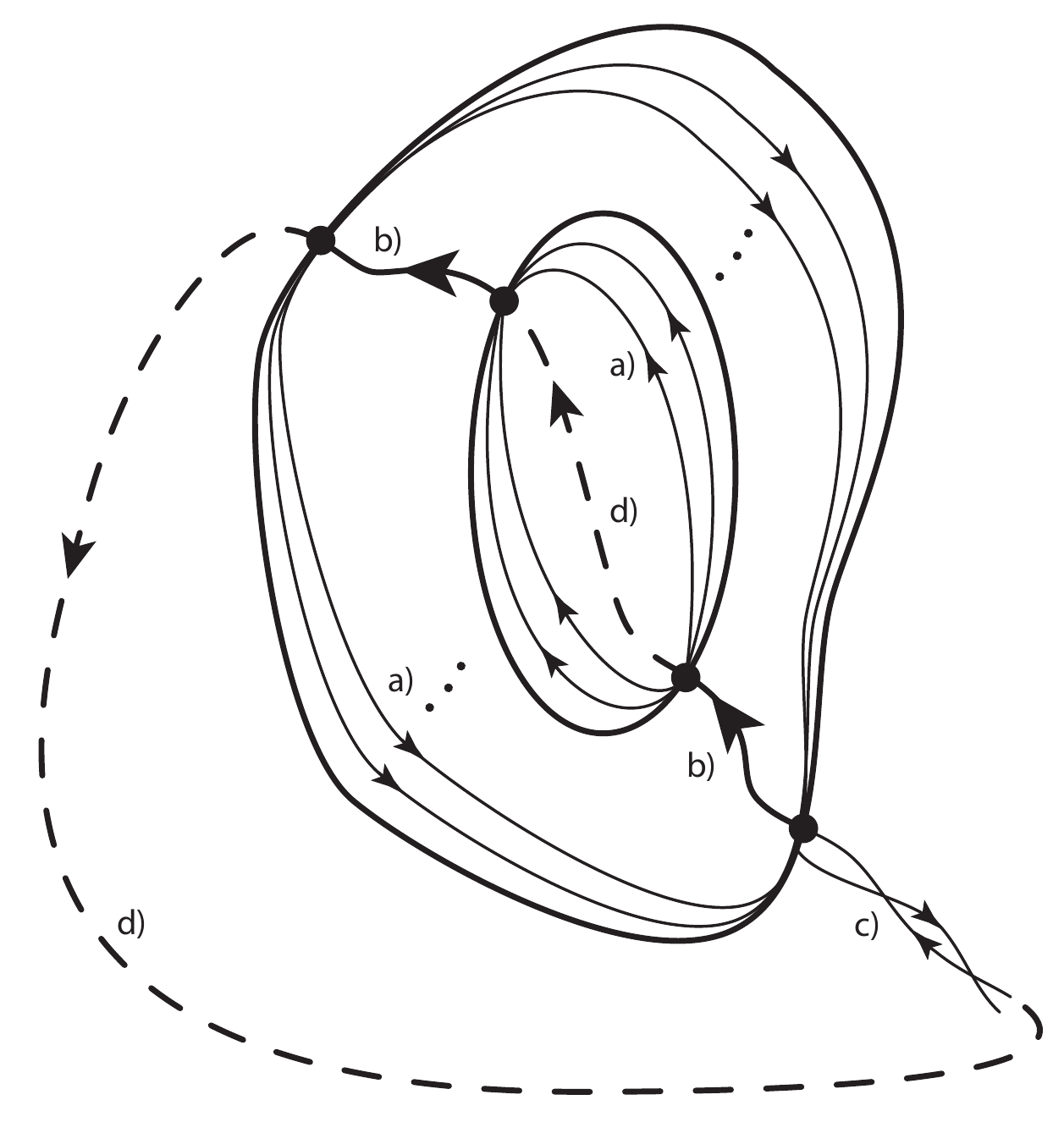}
\caption { An illustration of generalized axial windings, used to connect all equipotential contours of the Target and Return fields into a continuous helical winding.  a) axial currents flow from the maximum to minimum of $\Phi_{return}-\Phi_{target}$ on the inner surface of the Return and back from maximum to minimum of $\Phi_{return}$ on the outer surface.  b) two Jumpers connect the axial currents to make a complete circuit.  The corresponding toroidal winding field is confined to the Return so it does not disturb the inner or outer specified fields.  c) the entire coil is energized from a twisted pair inserted into one Jumper on the outer surface.  d) the two Jumpers can be extended outside the Return to form a filament circuit with calculable potential $\Phi_{wire}$ with the same singularity as $\Phi_{axial}$, so that the latter can be solved from the difference using standard Neumann boundary conditions.}
\label{fig:wiring}
\end{figure}

The construction described above will work with little modification for any system with an axis of rotation; the only difference is that the construction of $\Phi_{return}$ will entail solving the Laplace problem with Neumann boundary conditions in a more complicated geometry.  However, it can also be generalized to any Target region of arbitrary shape and physically allowable magnetic field.  The first step is to follow the basic algorithm~\cite{CBC2020} to learn how to produce a designed field, by finding the scalar potentials $\Phi_{target}$ and $\Phi_{return}$.  Equal intervals of $\delta \Phi$ describe ribbon loops on the surfaces which carry equal current.  Though this doesn't yet specify how to make a coil, it approximates its form. The points (the four dots in Fig.~\ref{fig:wiring}) where $\delta \Phi$ is maximal and minimal on each surface are the places where the coil on a surface must start and end.  For fields of dipolar character in a suitable Target and Return, there will be just one maximal and minimal point on each surface.  

Now we need to choose a ``wiring diagram'': Jumper wires connecting the minimal point on the Target to the maximal point on the Return, and vice versa for the other pair of extrema, with the current supply interpolated in one of these paths via a twisted pair.  At this point we also decide how large a current $I$ to use relative to the size of the field $H_0$ to be constructed (which determines the width of the ribbons, just as in the case of a solenoid). Assuming that the Return wraps around the Target, choose a branch surface whose boundary edge includes the wires connecting the Target to the Return, and traverses across the surfaces of the Target and the Return so that the surface interrupts any path through the Return that encloses the Target. Treating this as a special kind of boundary, the Return is now singly connected.

As in the case of the spherical electromagnet, we now construct a second scalar potential $\Phi_{axial}$ in the Return that has has vanishing normal derivative at the surface of the Target and the Return, and a discontinuity of magnitude $I$ across the branch surface.  This is not a standard Neumann boundary condition, though perfectly well defined.  

(Here's a way to turn it into a standard Neumann boundary value problem, as illustrated in Fig~\ref{fig:wiring}:  choose a wire circuit that carries the chosen current, such that it incorporates the two Jumpers, connected by wires which lie outside the Return and inside the Target. The Biot-Savart integral for the magnetic field of this circuit corresponds to a scalar potential $\Phi_{wire}(\vec r)=I\Omega/4\pi$ where $\Omega$ is the solid angle subtended by the circuit viewed from the point $\vec r$, and has a branch surface which can be chosen to coincide with that of $\Phi_{axial}$.  The difference between $\Phi_{axial}$ and $\Phi_{wire}$ is a solution to Laplace's problem inside the Return, and with normal derivative which is the difference between those of the two potentials.   The former is specified by the field in the Target and the latter is calculable, so this is a standard Neumann boundary value problem in a singly connected region.)  

As above we construct $\Phi_{chiral} = \Phi_{return} + \Phi_{axial}$ in the Return.   The difference $\Phi_{chiral} - \Phi_{target}$ is constant along a path that slices the surface of the Target into one long ribbon that defines the appropriate winding; similarly, $\Phi_{chiral}$ is constant along a path on the outer surface of the return that defines the winding there.  

The wires connecting the Target to the outer surface of the Return should join the extrema of the relevant potential differences, but the positioning of these has already been  determined earlier in the algorithm.  We claim that the wire positions continue to be appropriate:  the wires themselves are singular points of the fields where $\Phi_{chiral}$ takes on different values when the wire is approached along different paths.   Then the ``perturbation'' due to the tangential fields will automatically be absorbed in the determination of the surface windings.  

The axial potential $\Phi_{axial}$ produces an additional magnetic "winding field".  In the specific and general cases described above, this was a toroidal field circling about the Jumpers.  The boundary conditions were constructed to confine the winding field completely inside the Return, so that it didn't perturb either the Target or external fields.

The algorithm describes how to make a perfect electromagnet in the limit of a continuous surface current distribution when $I\to 0$.  The effect of replacing this with discrete windings is exponentially localized near the surfaces, with a healing length whose scale is set by the winding spacing~\cite{CBC2020}. This is largest near the wire connection points where the current density vanishes when the surfaces are smooth (this can be readily seen in the case of the spherical magnet), and has the consequence that the largest discreteness error occurs near the poles~\cite{Nouri}.  We propose that the Target and Return be deformed near the wire attachments, to make them somewhat conical about the wire (the Target becomes a lemon, and the Return has the shape of an apple).  The linear (rather than quadratic) variation of the potential near the connection point implies that the width of the spiralling ribbon will not grow so much in that limit, thus decreasing the healing length and making the field in the Target closer to the design field.  

In summary, this paper provides a general method to convert the series of disjoint equipotential contour loops described in \cite{CBC2020}, each carrying current $I$, into a continuous helical winding through use of an auxiliary axial potential $\Phi_{axial}$ representing the current $I$ flowing from the lowest to the highest equipotential along each surface of the Return.  This method is very general in the sense that the choice of wiring circuit, current, and equipotential constants classifies all possible helical windings which approximate the specified Target and external field with a discrete wire or trace winding.

\section{Acknowledgments}
This work is supported in part by the U.S. Department of Energy, Office of Nuclear Physics under contracts DE-SC0008107 and DE-SC0014622, and by the National Science Foundation under award number PHY-0855584.

\end{document}